\documentclass[12pt]{article} 
\catcode`@=11
\@addtoreset{equation}{section}
\begin{document}
\vskip 50pt
\begin{center}
{\Large \bf  Solitary Optical pulse propagation in fused fibre coupler-- Effect of Raman Scattering and switching .}
\end{center}
\vskip 20pt
\begin{center}
{\bf {  Basanti Mandal and A.Roy Chowdhury\footnote{E-mail: arcphy@cal2.vsnl.net.in}}\\ 
                           High Energy Physics Division \\      
                             Department Of Physics     \\           
                             Jadavpur University        \\     
                              Kolkata - 700032\\
                                    India \\}
\end{center}
\vskip 20pt
[    We analyse the motion of solitary type optical pulse in a fused optical fibre coupler when inter channel raman scattering (IRS) is taken into account and same type of dispersion management is used for both the core. The paper is divided into two sections . In the first half we utilise a moment method to study the variations of  the different parameters of the Gaussian pulse, such as chirp, frequency, energy etc. while in the second half assuming a general form of the stationary pulse we discuss the switching phenomena due to the coupling. It is shown that lauching a pulse in one arm  of the coupler can generate pulses in both the arms, but the patterns differs in a significant manner when the strength of IRS gets changed.   ]
\newpage
\section{Introduction}
Nonlinear directional couplers are now a days the building blocks of all optical processing. They are now subject of intense theoretical and experimental research[1]. The use of soliton like pulse in conjuction with these couplers has brought a revolution in the field of nonlinear optics[2]. Most directional couplers are usually twin core coupler with a parmanent coupling between the cores[3]. On the other hand interchannel Raman scattering is an important phenomenon which plays a decisive role in the formation and propagation of a pulse in a fibre[4]. So, in this communication we  have considered the mutual effect of intrapulse raman scattering (IRS) and the coupling on the propagation characteristics of the pulse in a fused coupler. These types of couplers  are now a days very important as they can be used for switching operation[5]. Here we have observed that a very weak pulse launched in one arm of the coupler can grow considerably due to the coupling[6]. A similar event takes place in case of conversion, compression and splitting of pulses.
\par  The paper is organised as follows. In the first section we derive the equations governing the evolution of the pulse parameters such as energy, width, centre, chirp etc. by following a moment method[6a]. Because of the presence of IRS, a variational procedure depending on a Lagrangian cannot be used[7]. These are then numerically solved and these parameters variations are studied with respect to the coupling strength and the magnitude of IRS.
\par In the second half of the present communication we have considered a general form of the stationary pulse and studied the effect of the coupling in the generation of pulse in both the arms when there is only one pulse initially in one arm. It is observed that the  presence of IRS strongly effects such switching pattern.

\section{Gaussian pulse parameter evolution: }
The equation describing the propagation of the nonlinear waves in a fused coupler is written as,
 $$\frac{\partial {u}}{\partial{z}} +\frac{i\beta _2}{ 2} \frac{\partial ^2{u}}{\partial {t ^2}} +iKv=i\gamma \left(\mid{u}\mid^2u-T_Ru\frac{\partial }{\partial{t} }\mid{u}\mid^2\right) \eqno{(1)}$$ 
 $$\frac{\partial {v}}{\partial{z}} +\frac{i\beta _2}{ 2} \frac{\partial ^2{v}}{\partial {t ^2}} +iKu=i\gamma \left(\mid{v}\mid^2v-T_Rv\frac{\partial }{\partial{t} }\mid{v}\mid^2\right) \eqno{(2)}$$ 
where $u(z,t)$ and $v(z,t)$ represent the wave forms in the two arms of the coupler, $\beta _2(z)$ is the dispersion management whose explicit forms are given in equations (20) and (21), $T_R$ is the coefficient of interchannel Raman scattering and K is the coupling constant. This coupled set is difficult to solve but can be studied numerically. One of the most popular method is to assume an initial profile of the incomming wave form (outgoing) and to study the variations of the physical parameters of the pulse propagating along the fibre. Unfortunately, the presence of the IRS term does not allow a suitable Lagrangian for the system. So the standard variational technique can not be implemented and we take recourse to the moment method. The basic idea is to treat the optical pulse like a particle whose energy (E), position (T), and frequency shift $(\Omega) $  (from the original carrier frequency) are defined as
$$E_\psi =\int_{-\infty } ^{\infty }\mid{\psi }\mid^2 dt \hskip85 pt  \eqno{(3)}$$
$$T_\psi =\frac{1}{E }\int_{-\infty } ^{\infty }t\mid{\psi }\mid^2 dt \hskip60 pt  \eqno{(4)}$$
$$\Omega_\psi =\frac{i}{2E }\int_{-\infty } ^{\infty }\left(\psi ^*\frac{\partial{\psi } }{\partial{t} }-\psi \frac{\partial{\psi ^*} }{\partial{t} }\right)dt\eqno{(5)}$$
The root mean square (RMS) width of the pulse is,
$${\sigma_\psi } ^2=\frac{1}{E }\int_{-\infty } ^{\infty }(t-T)^2\mid{\psi }\mid^2 dt\eqno{(6)}$$
The chirp of the pulse is given as,
$${\tilde C}_\psi =\frac{i}{2E }\int_{-\infty } ^{\infty }(t-T)\left(\psi ^*\frac{\partial{\psi } }{\partial{t} }-
\psi \frac{\partial{\psi ^*} }{\partial{t} }\right)dt\eqno{(7)}$$
where $\psi $ stands for either $u(z,t)$ or $v(z,t)$. The initial profile of the pulse u,v are assumed to be Gaussian-type,
$$\psi (z,t)=\sqrt\frac{ E_\psi }{\pi\tau_\psi   }exp\left[i \phi_\psi  -i \Omega_\psi  (t-T_\psi  ) -(1+iC_\psi  )\frac{(t-T_\psi  )^2}{2\tau_\psi   ^2}\right] \eqno{(8)}$$
Differentiating (3) to (7) and using (1), (2) and (8) we get the following system of nonlinear coupled evolution equations ,
$$\frac{dE_u }{dz }=-2K\; P_1(E,\tau )\;sin(\phi _1-\phi _2)\;e^{-\frac{1}{2}(T_u-T_v)^2}\hskip 145pt \eqno{(9)}$$
$$\frac{dT_u }{dz }=\beta _2\Omega _u-\frac{ 2K}{E_u }P_3(E,\tau )\;sin(\phi _1-\phi _2)\;\left [2T_u\tau _v^2+(T_u+T_v)\tau _u^2\right ]e^{-\frac{1}{2}(T_u-T_v)^2} \eqno{(10)}$$
$$\frac{d\Omega _u }{dz }=-\frac{ \gamma T_RE_u}{ \pi \sqrt{2\pi}\tau _u^3}+\frac{2K }{E_u }P_3(E,\tau )\;e^{-\frac{1}{2}(T_u-T_v)^2}\times\;\left [(T_v-T_u)cos(\phi _1-\phi _2)\right.\hskip 43pt $$$$\left.+\left\{(\tau _u^2+\tau _v^2) (\Omega _u-\Omega _v)-(T_u-T_v)C_v\right\}sin(\phi _1-\phi _2) \right]\eqno{(11)}$$
$$\frac{d\tau _u }{dz }=\frac{\beta _2C_u }{\tau _u }+\frac{K }{2E_u\tau _u }P_5(E,\tau )\;sin(\phi _1-\phi _2)\;e^{-\frac{1}{2}(T_u-T_v)^2}\times \hskip 112pt$$$$\left[(\tau _u^2+\tau _v^2)\tau_u^2+4 (\tau _u^2+\tau _v^2)\tau _v^2-4\tau _u^2(T_u-T_v)^2\right] \eqno{(12)}$$
\newpage
$$\frac{dC_u }{dz }=-\beta _2\Omega _u^2+\frac{\beta _2 }{2\sqrt{\pi}\tau _u^2 }(1+C_u^2+2\Omega _u^2\tau _u^2)+\frac{\gamma E_u }{\tau _u(2\pi)^{3/2} } +\frac{K }{E_u }P_5(E,\tau )\hskip 40pt$$$$\times e^{-\frac{1}{2}(T_u-T_v)^2} \left[cos(\phi _1-\phi _2)\left\{\tau _v^4-\tau _u^4-2(\tau _u^2T_v+\tau _v^2T_u)(T_u-T_v)\right\}\right.$$ 
$$+sin(\phi _1-\phi _2)\left\{C_u(\tau _v^4-\tau _u^4)-2\Omega _u(\tau _u^2+\tau _v^2)\left(2T_u\tau _v^2+(T_u+T_v)\tau _u^2\right)\right.$$ $$\left.\left.-2(\tau _u^2T_v+\tau _v^2T_u)\left(C_v(T_u-T_v)+\Omega _v(\tau _u^2+\tau _v^2)\right)\right\}\right]\eqno{(13)}$$
$$\frac{dE_v }{dz }=2K\; P_1(E,\tau )\;sin(\phi _1-\phi _2)\;e^{-\frac{1}{2}(T_u-T_v)^2} \hskip 150pt \eqno{(14)}$$
$$\frac{dT_v }{dz }=\beta _2\Omega _v+\frac{ 2K}{E_v }P_3(E,\tau )\;sin(\phi _1-\phi _2)\;\left [2T_v\tau _u^2+(T_u+T_v)\tau _v^2\right ]e^{-\frac{1}{2}(T_u-T_v)^2} \eqno{(15)}$$
$$\frac{d\Omega _v }{dz }=-\frac{ \gamma T_RE_v}{ \pi \sqrt{2\pi}\tau _v^3}-\frac{2K }{E_v }P_3(E,\tau )\;e^{-\frac{1}{2}(T_u-T_v)^2}\times\; \left [(T_v-T_u)cos(\phi _1-\phi _2)\right.\hskip 40pt$$$$\left.+\left\{(\tau _u^2+\tau _v^2) (\Omega _v-\Omega _u)-(T_v-T_u)C_v\right\}sin(\phi _1-\phi _2) \right]\eqno{(16)}$$
$$\frac{d\tau _v }{dz }=\frac{\beta _2C_v }{\tau _v }-\frac{K }{2E_v\tau _v }P_5(E,\tau )\;sin(\phi _1-\phi _2)\;e^{-\frac{1}{2}(T_u-T_v)^2}\times \hskip 110pt$$$$\left[(\tau _u^2+\tau _v^2)\tau _v^2+4(\tau _u^2+\tau _v^2)^2\tau _u^2-4\tau _v^2(T_u-T_v)^2\right] \eqno{(17)}$$
$$\frac{dC_v }{dz }=-\beta _2\Omega _v^2+\frac{\beta _2 }{2\sqrt{\pi}\tau _v^2 }(1+C_v^2+2\Omega _v^2\tau _v^2)+\frac{\gamma E_v }{\tau _v(2\pi)^{3/2} } +\frac{K }{E_v }P_5(E,\tau )\;\hskip 50pt$$$$\times e^{-\frac{1}{2}(T_u-T_v)^2} \left[cos(\phi _1-\phi _2)\left\{\tau _u^4-\tau _v^4-2(\tau _u^2T_v+\tau _v^2T_u)(T_v-T_u)\right\}\right.$$$$-sin(\phi _1-\phi _2)\left\{C_v(\tau _u^4-\tau _v^4)-2\Omega _v(\tau _u^2+\tau _v^2)\left(2T_v\tau _u^2+(T_u+T_v)\tau _v^2\right)\right.$$ $$\left. \left.-2(\tau _u^2T_v+\tau _v^2T_u)\left(C_u(T_v-T_u)+\Omega _u(\tau _u^2+\tau _v^2)\right)\right\}\right]\eqno{(18)}$$
This set of ten ordinary differential equations is then integrated by the Runge-Kutta-Felberg algorithm. In the above equations,
$$P_i=\sqrt{\frac{ 2E_uE_v\tau _u\tau _v}{\pi(\tau _u^2+\tau _v^2)^i }}\eqno{(19)}$$
The above ordinary differential equations can now be used to simulate two different physical phenomena. One is the coupling between the two modes of propagation on the two cores and the other is the effect of Raman interchannel scattering and not the least their mutual influence on each other. We also show the initial and final form of $|u|^2$ and $|v|^2$ to study the effect of switching in the coupler. Before proceding to the actual numerical simulation, we comment on the type of dispersion management used,
 \newpage Type (a) $$\beta _2(z)=\delta _a +\frac{1}{z_a}\bigtriangleup (\zeta ) \eqno(20)$$
\par  with dispersion map,
$$\bigtriangleup (\zeta )=\left \{\begin{array} {c}\bigtriangleup _1\; \;  ;\;\;\;\; 0\leq \mid \zeta  \mid < \frac{\theta  }{2}\\ \bigtriangleup_2 \; \;  ;\;\;\;\;  \frac{\theta }{2}\leq \mid \zeta  \mid <\frac{1}{2} \end {array}\right\} $$
where 
$$ \bigtriangleup _1=\frac{2s}{\theta }\; , \;\;\bigtriangleup _2= -\frac{2s}{1-\theta }\;\;\; $$
and map strength, $$s=\frac{\theta \bigtriangleup _1- (1-\theta )\bigtriangleup_2}{4}$$
\par Type (b) $$ \beta _2(z)=\frac{\delta sin2z }{(1+\delta sin^2z)^2 } \eqno(21)$$
The type (a) was used by many workers[8] with an eye to the actual construction of the anomalous and normal section of a optical fibre, with specified gap of amplification. On the other hand the second type of dispersion management was used by  Serkin  etal[9]. for explicit analytic construction of soliton through inverse scattering to be used as a model of optical pulse.
\par The results of our numerical simulation are clearly displayed in figures. To start with                  consider figure(1) where the variation of E(energy), $\Omega (RIFS)$, $\tau$(width) and C(chirp) are plotted with respect to distance travelled by the pulse. In each of the figure we have shown the effect of variation of the coupling  constant K keeping all other parameters fixed. The type of dispersion management used is given in equation (21), with $\delta= - 0.9$. On the other hand in figure (2) the same type of dispersion management with different value of $\delta $(=0.9) is shown. In both the cases the value of $T_R$ is the same. On the other hand due to the usual type of dispersion map given in equation (20) , the variation of E, C, RIFS, $\tau $ are exhibited in figures (3), (4), (5) for different values of K. One may note that the variation of energy (E) in figure (5) is shown that while it decreases in one core, in the second one it grows. The same phenomenon is clearly visible in figure (6) where the growth in one and decay in the other are exhibited. So this kind of phenomenon is actually a switching action for which such couplers are constructed.The final shape of output pulse in both the cases are exhibited in figures(6) and (7).  The corresponding parameter values are given in the figure caption. It may be noted that even if we keep the amplitude of the input pulse in one of the channel very small, still there is a significant display of the pulses in both the cores. One may observe that it is not possible to set one of pulses exactly equal to zero due to  some mathematical inconsistencies It may be added that all these analysis can easily be repeated if the initial profile of the pulse is assumed to be of such type, rather than Gaussian are.
 \section{Switching Action:}
 The switching action can be further studied if we neglect the dispersion management and consider the simplified model which exhibits translational invariance. Assuming propagating pulse of the form$$ u=f(z-\omega t) \;\;\; and \;\; v=g(z-\omega t) $$ we get
$$\frac{du}{d\xi} +i\omega ^2\frac{\beta _2}{ 2} \frac{d^2u}{d\xi ^2} +iKv=i\gamma \left(\mid{u}\mid^2u+T_R\;\omega\; u\frac{d }{d\xi}(\mid{u}\mid^2)\right) \eqno{(22)}$$ 
 $$\frac{dv}{{d\xi }} +i\omega ^2\frac{\beta _2}{ 2} \frac{d^2v}{d\xi  ^2} +iKu=i\gamma \left(\mid{v}\mid^2v-T_R\;\omega\; v\frac{d }{d\xi }(\mid{v}\mid^2)\right) \eqno{(23)}\hskip 5pt$$ 
Here $\beta _2$ and $\gamma $ all are constants. Writing $u=u_1+iu_2$ and $v=v_1+iv_2$ and separating real and imaginary parts we get four ordinary differential equations of second order. Assuming that  at the launch condition v = 0, and $u = u_0$ , we have integrated the system by Runge-Kutta, ensuring the boundary condition needed for a solitary pulse. The shape of the pulse as it imerges is shown in the figure (8) for various coupling constan K. Note that here the input pulse is kept zero in one arm, but in the output side, it can be seen in both the arms.
.
\section{Conclusion:}
Here in this note we have analyzed a model of a fused coupler, where both the dispersion management and IRS are assumed to be present. It is very clear that the type of dispersion management significantly effects the variation of the parameters, on the other hand the IRS causes some irregularities in their variation.
\section{Acknowledgement} One of the authors (B.M) is grateful to U.G.C. (Govt. of India) for junior research fellowship. 
\section{References:}
1. {M. Liu and P. Shum} - {\it  Optics Express. } {\bf\underline{11}} 116 (2003). \\ 
2. {G.P.Agarwal} - {\it Fibre Optic Communication System.}- John Wiley \& Sons, 
\par New York, 1997.\\
3. {J. Santhano, G.P.Agarwal}- {\it Optics Communication.} - {\bf\underline{222}} (2003) 413.\\ 
4. {V. Serkin and A. Hasegawa} - {\it Physics / 0002027.}\\
5. {P. L. Chu, V. Kivshar and B. A. Malomed, G. P. Peng, M. L. Q. Teixeiro}-
\par {\it   J. Opt. Soc. Am. B.} {\bf\underline{12}}  898 (1995). \\
6. {A. Kumar and A. Kumar } - {\it   Optics Communication.} {\bf\underline{150}} (1998) 91. 
\par {J. J. G. Ripoll and V. M. P. Garcia} - {\it Patt-Sol.}/ 9904006. (preprint) .\\
7. {A. B. Moubissi, K. Nakkeeran, P. T. Dinda and T. C. Kofane.} - 
\par {\it J. Phys.A.} {\bf\underline{24}} (2001) 129.\\ 
8. {T. Hirooka and A. Hasegawa} - {\it Optics Letters. }{\bf\underline{23}} (1998) 768.
 \par {M. J. Ablowitz and G. Biondini} - {\it Opt. Lett. }{\bf\underline{23}} (1998) 1668.
\par {J.H.B. Nijhof, N.J.Doran, W.Forysiak, A.Berntson} - {\it Electron. Lett.}
\par {\bf\underline{34}} (2000) 481.\\
9. {V.N. Serkin, M. Matsumato, T.L. Belyaeva}-{\it Opt.Communications.}   {\bf\underline{196}}(2001)196 .
\par{V. Serkin and A. Hasegawa} - {\it IEEE. J. Quant. Electronics. }{\bf\underline{8}} (2002) 418.

\newpage
\section{Figure Caption}
Figure-1.    Variation of parameters, $E, \Omega, \tau$ and C  of u and v with respect to z taking $\delta=-0.9$where dotted(..) line stands  for k=-3, dash-dotted(-.)  line for k=3 and solid(-) line for k=0.9 and $T_R=0.696, \gamma=0.1994$\\
Figure-2.  Variation of  parameters, $E, \Omega, \tau$ and C of u and v with respect to z taking $\delta=0.9$where dotted(..) line stands  for k=-3, dash-dotted(-.)  line for k=3 and solid(-) line for k=0.9 and $T_R=0.696, \gamma=0.1994$\\
Figure-3. Variation of E and C of u and v with respect to  propagation distance when $T_R=0.5, \gamma= 0.1994$ and k=1.0\\
Figure-4.  Variation of E and C of u and v with respect to  propagation distance when $T_R=1.5, \gamma= 0.1994$ and k=2.0\\
Figure-5.  Variation of $\Omega and \tau$ of u and v with respect to  propagation distance when $T_R=0.696, \gamma= 0.1994$ and k=0.09\\
Figure-6.  Output pulses with respect to  propagation distance when $T_R=0.696, \gamma= 0.1994$ and k=0.09\\
Figure-7.  Output pulses with respect to  propagation distance when $T_R=0.696, \gamma= 0.1994 and k=0.09$ with different initial profile.\\
Figure-8a.  Output pulses with respect to  propagation distance with $\gamma=0.02, \beta_2=-0.02, \omega=0.9, T_R=1.6 $\\
Figure-8b.  Output pulses with respect to  propagation distance with $\gamma=0.02, \beta_2=-0.02, \omega=0.9, T_R=0.5 $\\
Figure-8a.  Output pulses with respect to  propagation distance with $\gamma=0.02, \beta_2=-0.02, \omega=0.9, T_R=3.0 $\\

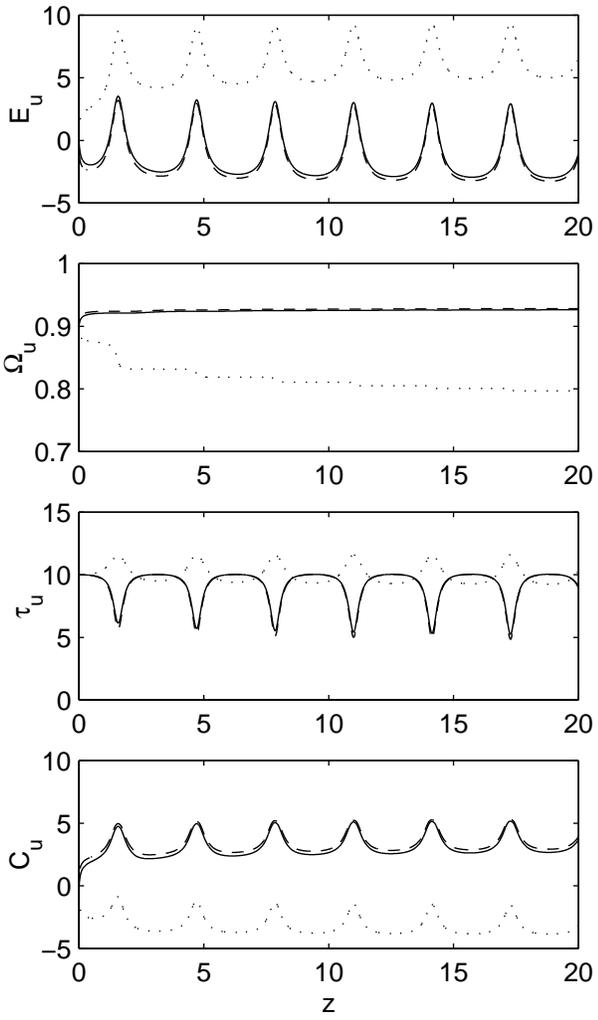
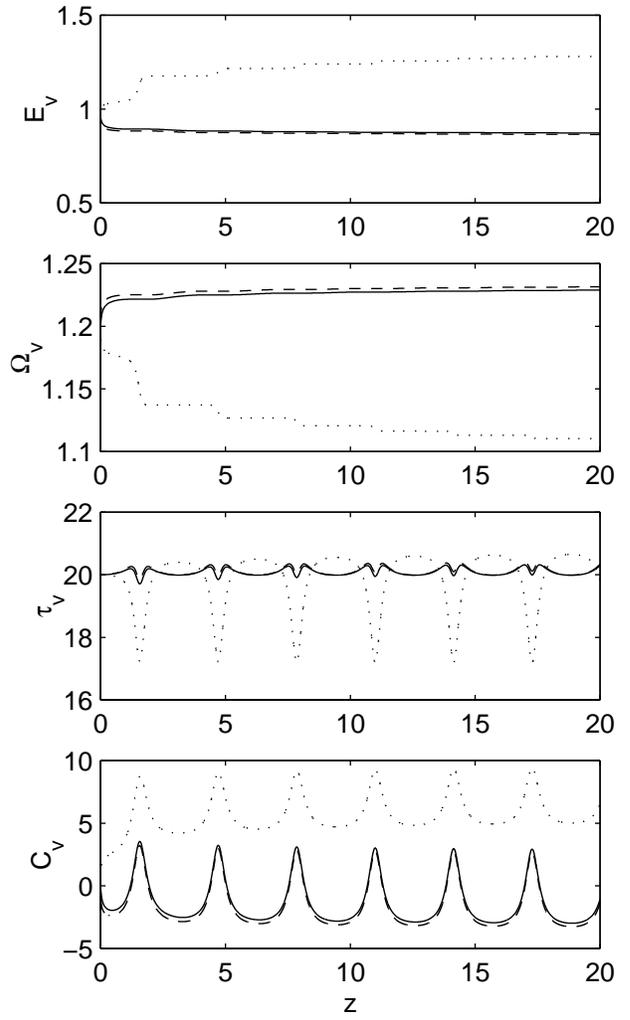
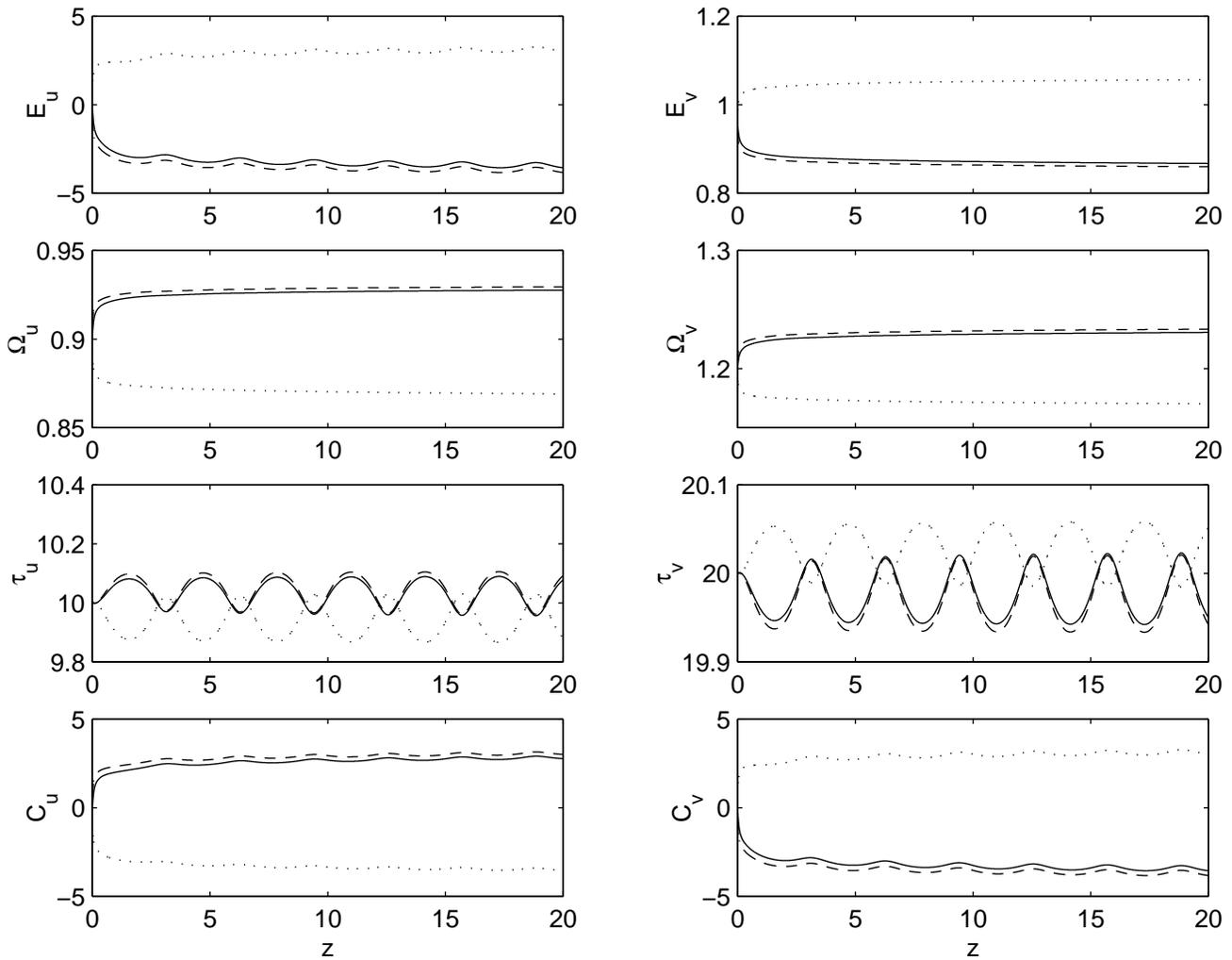
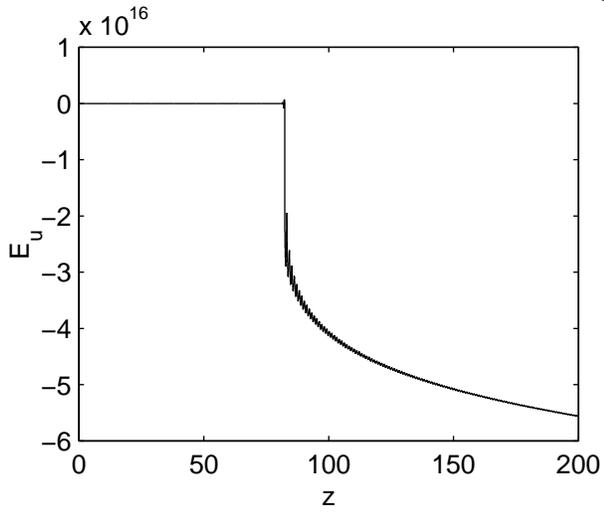
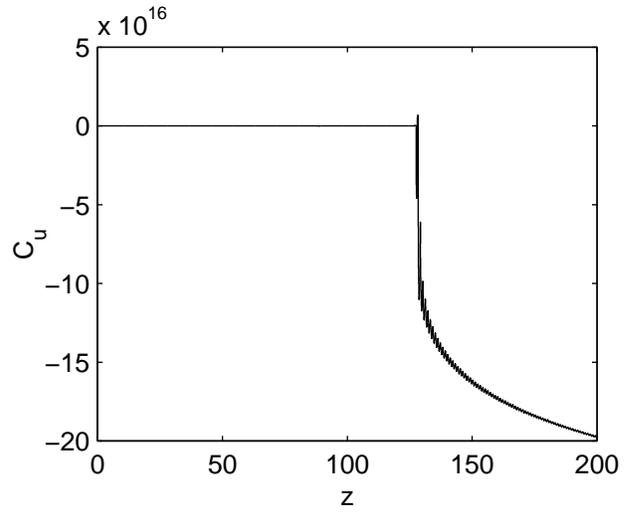
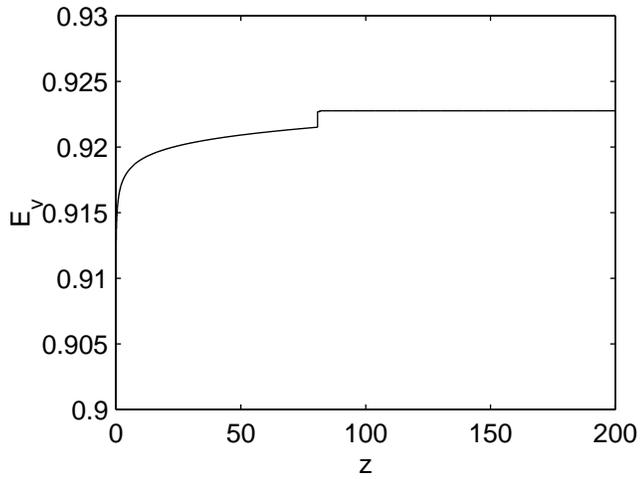
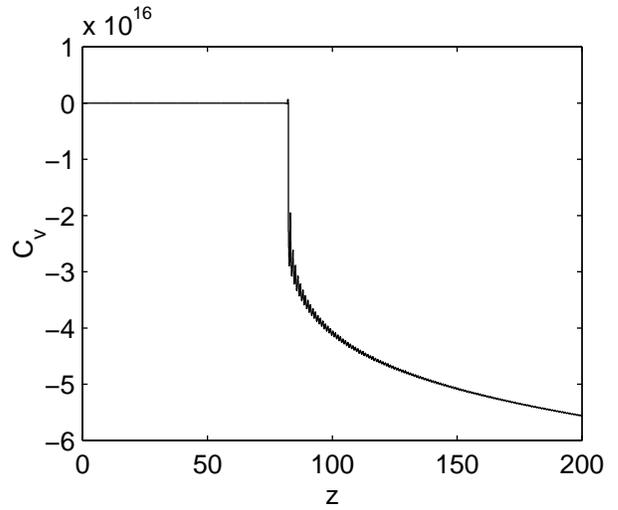
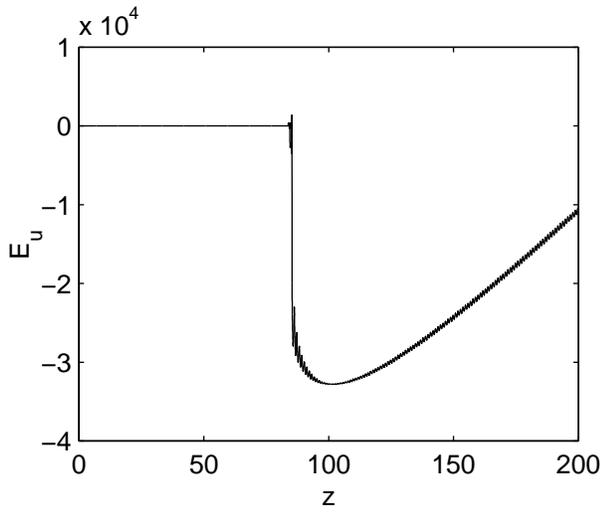
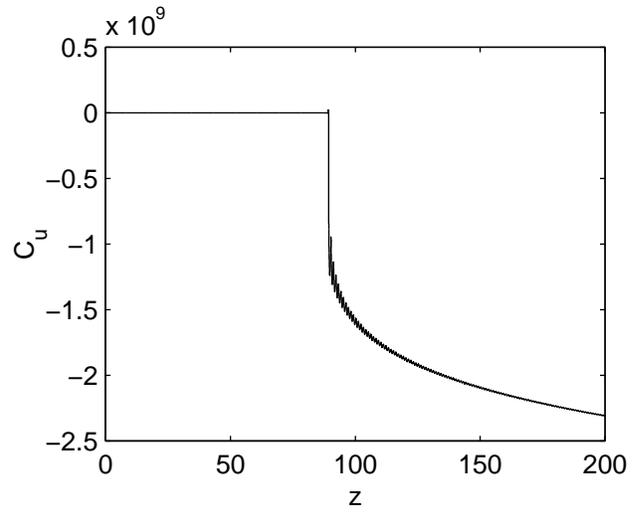
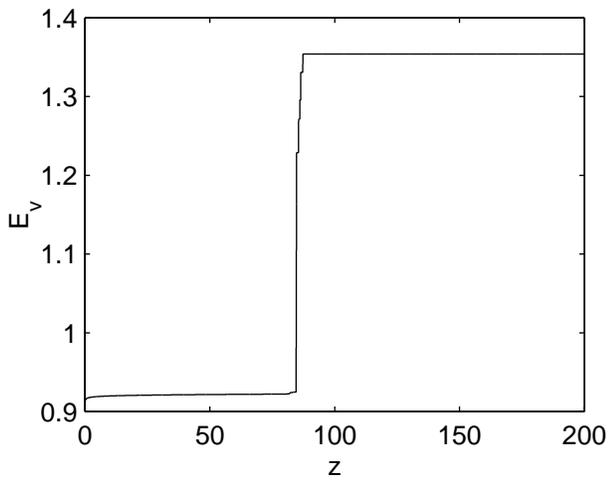
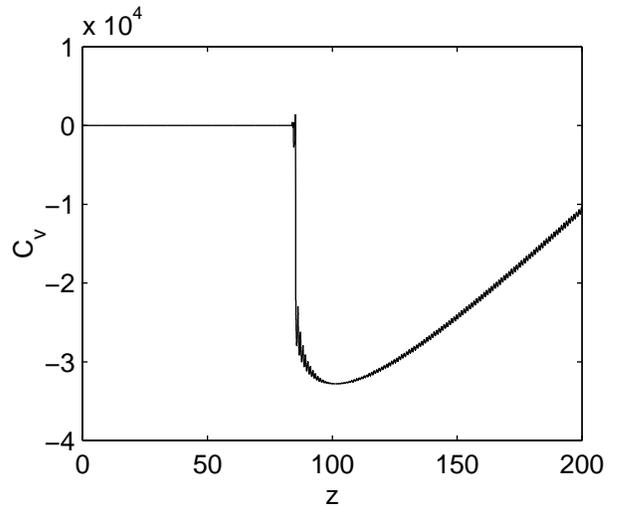
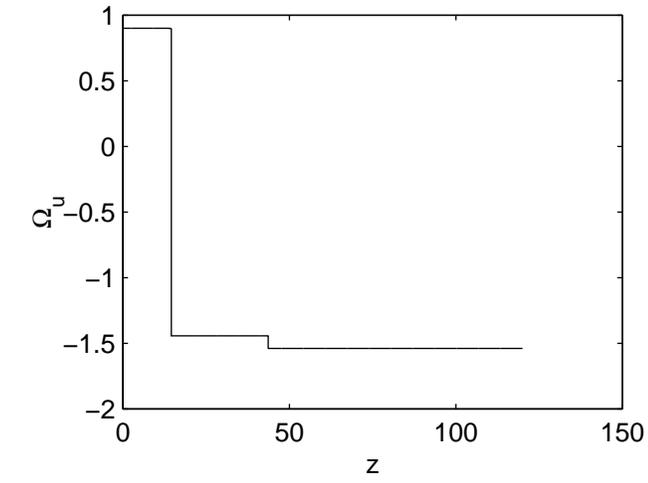
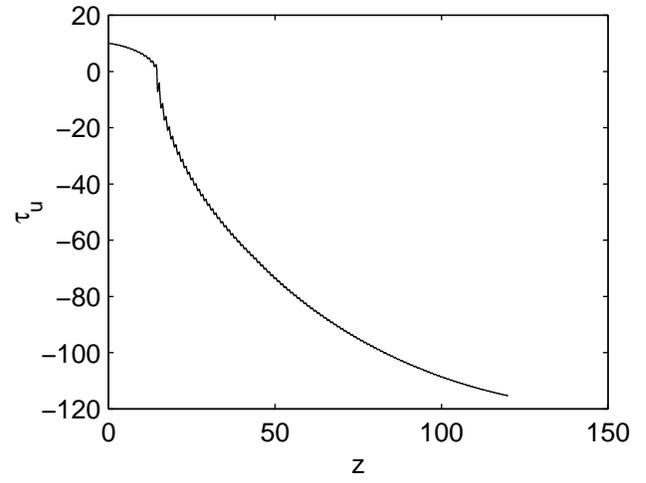
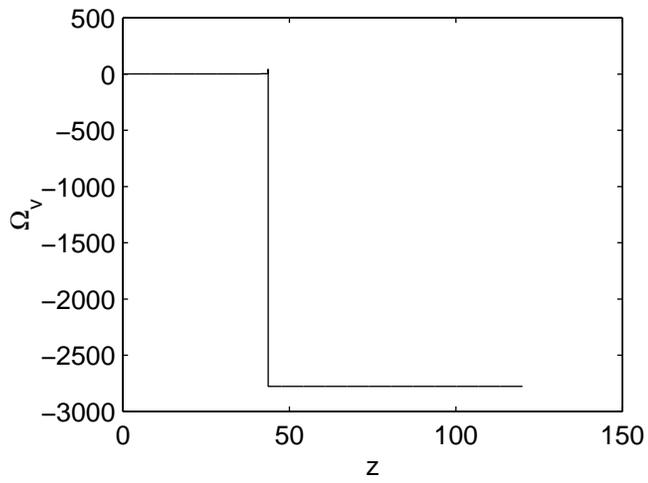
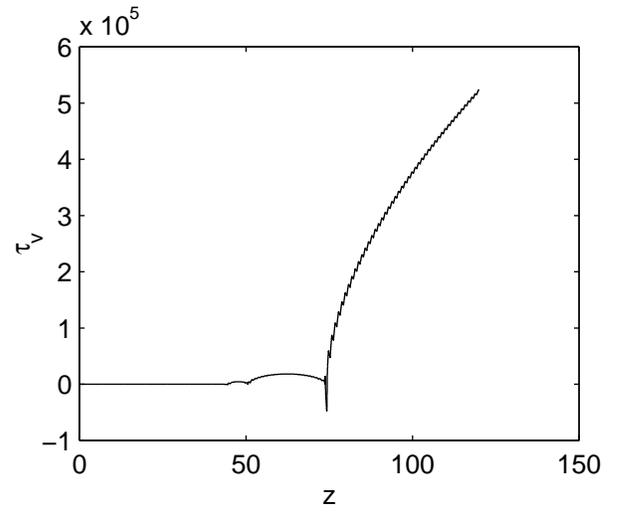
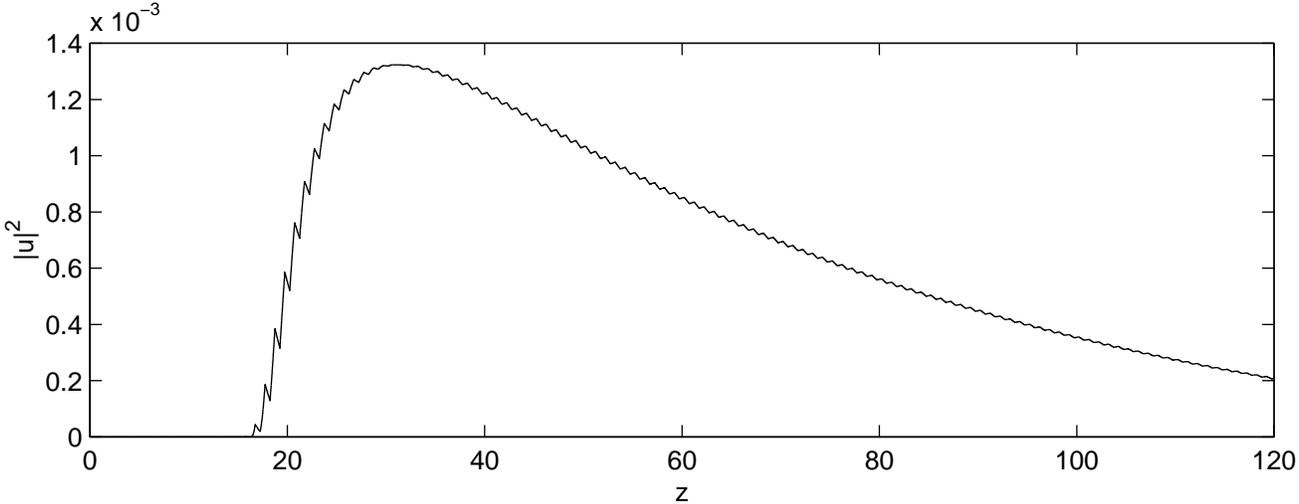
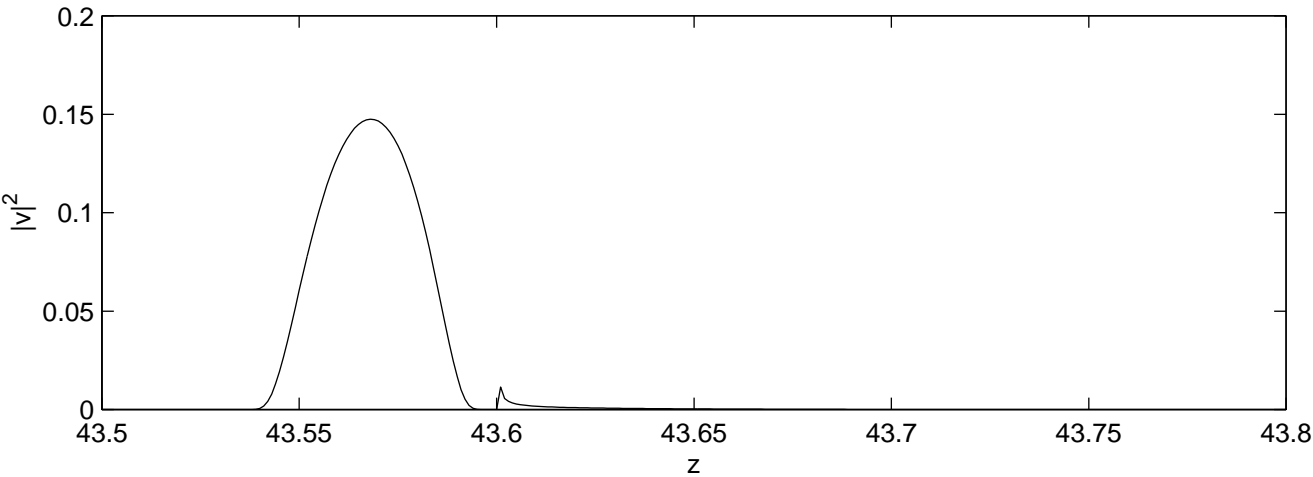
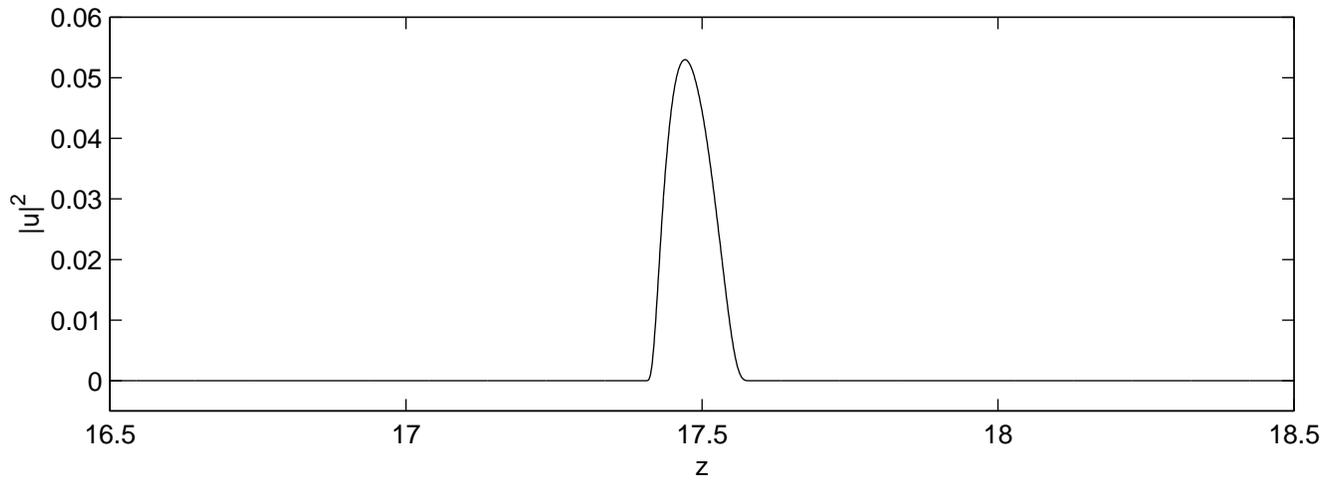
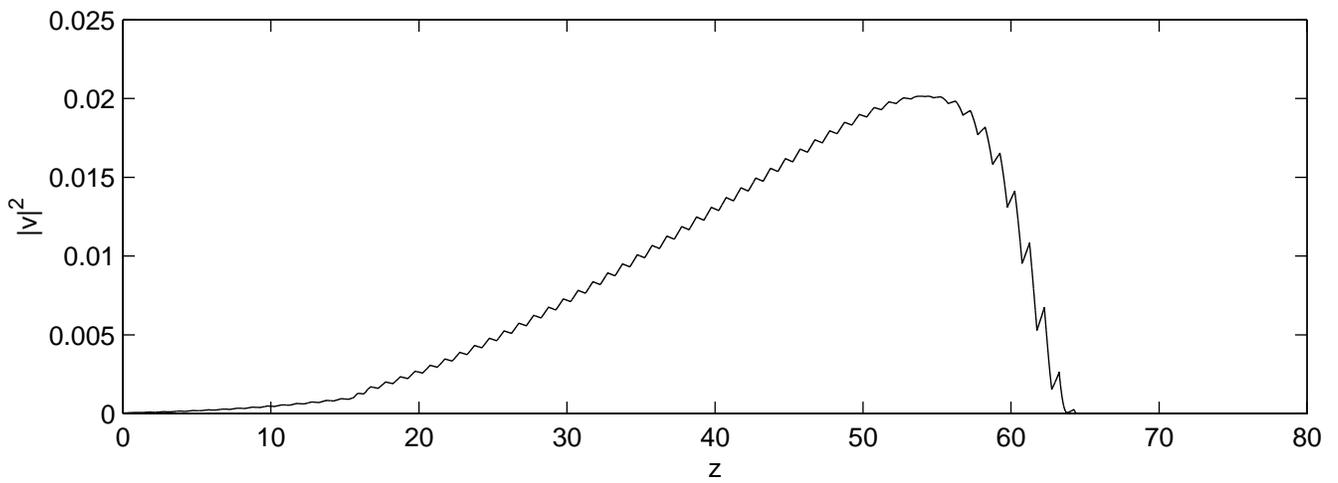
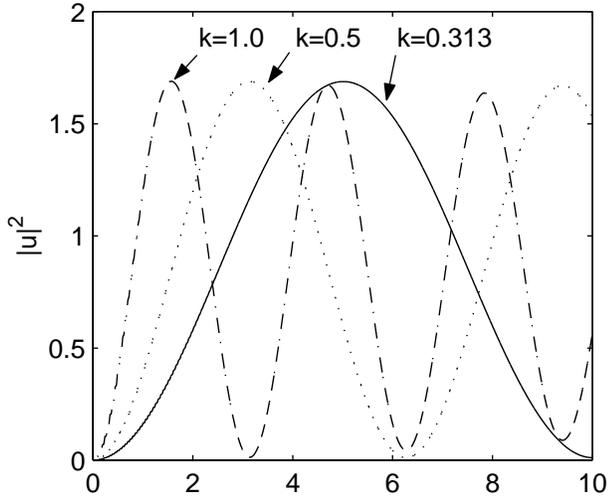
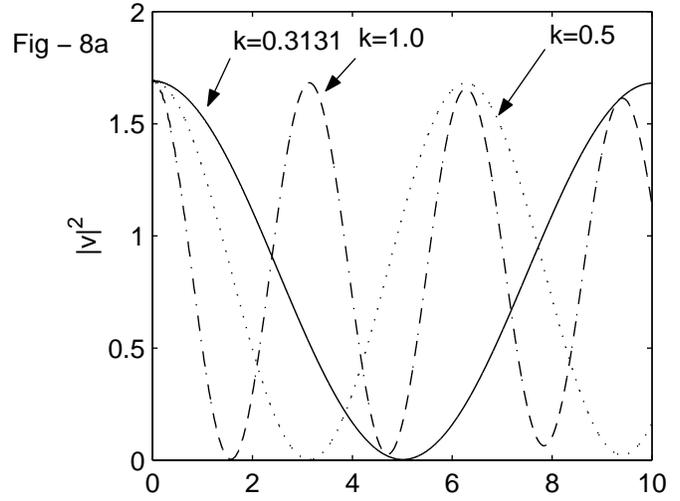
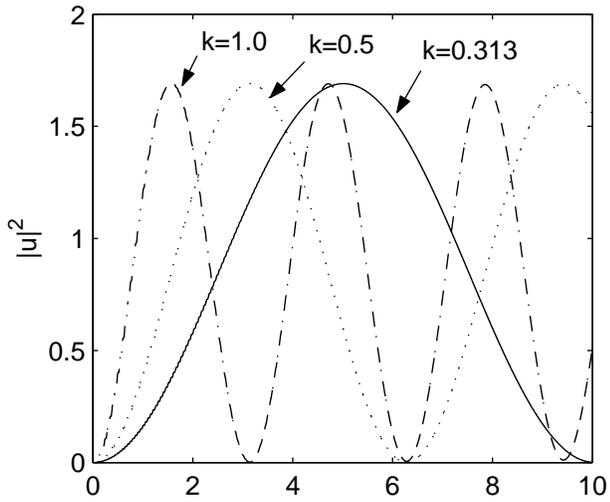
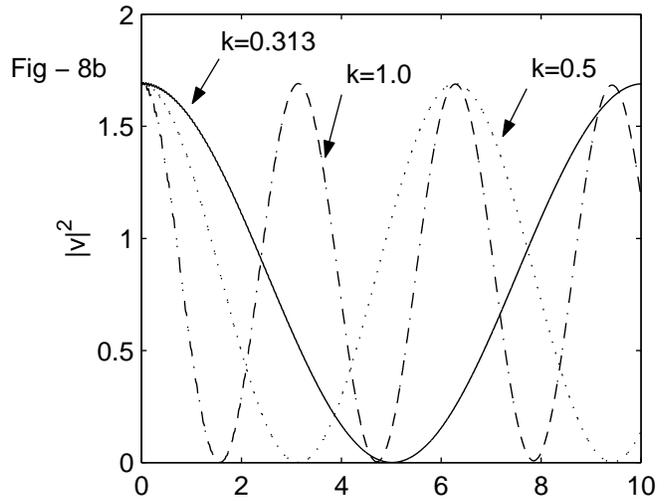
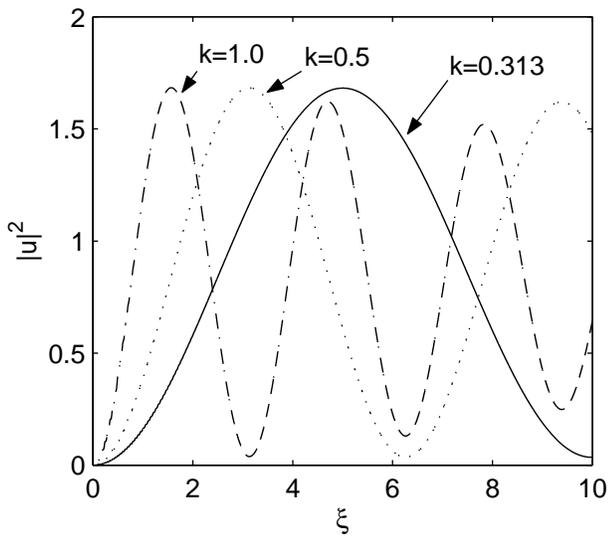
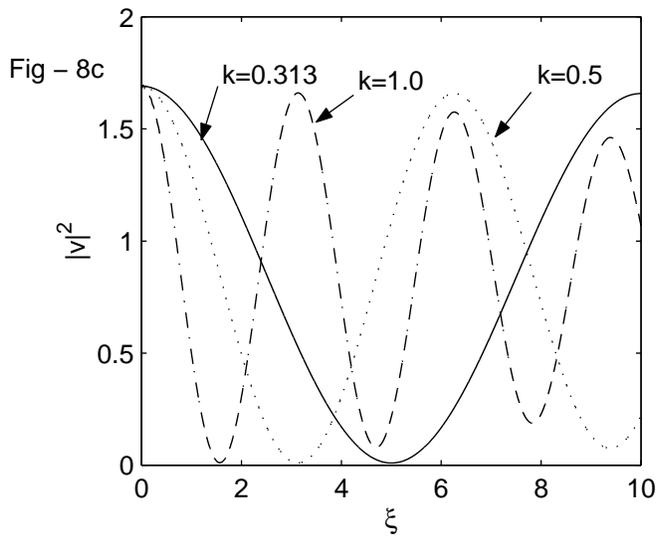
\end{document}